\documentstyle[aps,epsfig]{revtex}  
\begin{document}

\draft

\title{Quark scattering at QGP/2SC interface}  
\author{M. Sadzikowski\footnote{Foundation for Polish Science Fellow.}}   
\address{  Center for Theoretical Physics, Massachusetts Institute of Technology\\
           Cambridge, MA USA 02139 \\
           and\\
           Institute of Nuclear Physics, Radzikowskiego 152, 
             31-342 Krak\'ow, Poland}    
\maketitle    

\tighten    

\begin{abstract} We considered the phenomenon of the Andreev reflection of quarks at the interface
between the cold quark-gluon plasma phase and 2SC color superconductor.
\end{abstract}

\bigskip

%\begin{flushright}    
%Preprint number: MIT-CTP-3234
%\end{flushright}

\section{Introduction}    

The very interesting phenomenon arises at the junction between a   
normal-metal and a superconductor. The reflection of a conduction electron from the  
superconducting barrier,
first elaborated in the paper \cite{andreev}, is known in the literature as the Andreev reflection.  
It was shown that the excitation in the normal metal outgoing from the reflection point has the sign 
of the momentum,  the sign of the charge, and the sign of the effective mass opposite to that of the incoming  
electron  (hole quantum numbers). This peculiar prediction was checked by tunneling experiments and ballistic 
(direct)  observations \cite{exp}.    

In this letter we raise the interesting question of the interaction between the  
phase containing the free quarks (QGP) and the color-superconductor (CS). 
We shall show that the free quarks from the QGP phase falling at the CS barrier
are reflected in the similar fashion as was described by Andreev in \cite{andreev}.

In the next section we develop the Bogolubov - de Gennes equations for superconducting phase. 
The third section contains the detailed discussion of the Andreev reflection. 
In the last chapter we make some comments about the influence of the Andreev reflection on the phenomena
that can take place inside the Neutron Stars.  

\section{Bogolubov - de Gennes equations for 2SC phase}  

Our interest is the description of the phenomena that happens at the QGP/CS 
junction at moderate densities expected in the Neutron Stars. In this work we do not address the question of 
the gluon interaction with the CS phase thus we just concentrate on the fermionic degrees of freedom. 
The model of QGP is the gas of the free quarks. 
The quarks in QGP built the Fermi ladder, with the Fermi energy $\mu $ (of
order $400 - 500$ MeV in the Neutron Stars), where $\mu $ is a quark chemical potential.
In the case of the two light quarks, which we consider here, 
the CS phase creates the 2SC superconductor (for the review of the subject see \cite{raja1}).  
The essential physics of 2SC phase at moderate densities, is described by the effective four-fermion point 
interaction with the attractive pseudoscalar channel \cite{raja2}. This model leads to the Cooper instability 
of the Fermi sea which creates the new vacuum of condensed Cooper pairs. The physical excitations are fermionic 
quasiparticles separated from the vacuum by the energy gap $|\Delta |$. The energy gap was calculated in the 
mean-field approximation in the variety of models giving the values $40-140$ MeV which is about 10-30 per cent 
of the Fermi energy. 

The effective hamiltonian at the mean-field level takes the form:
\begin{equation}
H=\int d^3x\left\{\psi^{j\, \dagger}_{\alpha }\left( -i\vec{\alpha }\cdot\vec{\nabla }+m\gamma_0-\mu\right)\psi^{j}_\alpha +
\frac{\Delta }{2}\psi^{j\, \dagger}_{\alpha } (\tau_2)_{jk} t_{\alpha\beta } C\gamma_5 \psi^{k\, \dagger }_{\beta }-
\frac{\Delta^\ast }{2}\psi^{j}_{\alpha } (\tau_2)_{jk} t_{\alpha\beta } C\gamma_5 \psi^{k}_{\beta }+{\cal H}(\Delta ,m)\right\}
\end{equation}
where $m$ is a quark mass and $\psi^{j}_\beta $ is
the field operator of the quark. The greek indices describe the color and the latin indices
describe the isospin quantum numbers. The matrix $\tau_2$ is the antisymmetric Pauli matrix and $t$ is
one of the three antisymmetric Gell-Mann matrices. For our calculations we chose it arbitrarily as the  $\lambda_2$.
The gap parameter $\Delta (\vec{r})$ depends on the position. It vanishes inside the QGP phase and it takes
a nonzero value in the 2SC phase.
The equation of motion describing our system takes the form:
\begin{eqnarray}
\label{eom}
i\dot{\psi }^{j}_{\beta }=\left( -i\vec{\alpha }\cdot\vec{\nabla } +  m\gamma_0 - \mu\right)\psi^{j}_{\beta }
+ \Delta (\vec{r})  (\tau_2)^{j}_{k}t^{\sigma }_{\beta } C\gamma_5 \psi^{k\,\dagger\, T }_{\sigma } \\\nonumber
i\dot{\psi }^{j\,\dagger }_{\alpha }=-\vec{\nabla }\psi^{j\,\dagger }_{\alpha }\cdot i\vec{\alpha } - 
\psi^{j\,\dagger }_{\alpha }\left( m\gamma_0 - \mu\right) +
\Delta (\vec{r})^{\ast } \psi^{k\,T}_{\beta }(\tau_2)^{j}_{k}t^{\beta }_{\alpha }C\gamma_5
\end{eqnarray}
The equations (\ref{eom}) describe the interaction of the 6 ``families'' of quarks coupled through the
non-zero value of the gap parameter.  Using standard conventions (e.g. \cite{sad}),
one can find that the up, "red" quark couples
to the down "green" quark, the up "green" to the down "red", whereas the "blue" quarks remain free.  For
the first ``family'' the equations (\ref{eom}) read:
\begin{eqnarray}
\label{eom1}
i\dot{\psi }^{u}_{R}=\left( -i\vec{\alpha }\cdot\vec{\nabla } +  
m\gamma_0 - \mu\right)\psi^{u}_{R} - \Delta (\vec{r}) C\gamma_5\psi^{d\,\dagger\, T}_{G}\\\nonumber
i\dot{\psi }^{d\,\dagger }_{G}=-\vec{\nabla }\psi^{d\,\dagger }_{G}\vec{\alpha } 
-\psi^{d\,\dagger }_{G}\left(  m\gamma_0 - \mu\right) -
\Delta (\vec{r})^{\ast } \psi^{u\,T}_{R}C\gamma_5
\end{eqnarray}
where $u$ and $d$ are isospin and $R$ and $G$ color degrees of freedom.
The equations for the second ``family'' (up ``green'' and down ``red'' quarks) can be evaluate exactly 
the same way and the
"blue" quarks are not interesting in the first approximation. 
One can change the operator equations (\ref{eom1}) to c-number equations
by taking the expectation values $f \equiv\langle\Psi_0| \psi^{u}_{R} |\Psi_1\rangle $
and $g^\dagger\equiv \langle\Psi_0| \psi^{d\,\dagger }_{G} |\Psi_1\rangle $, where $|\Psi_0\rangle\, ,|\Psi_1\rangle $
are ground and excited state respectively. In the QGP phase ($|\Delta | =0$) the fields 
$f$ and $g^\dagger $ describe a
``particle'' and a ``hole''. Let us now find the quasiparticle wavefunctions
in the case of $\Delta = const$. It is convenient to decompose the fermi fields in the
plane wave basis.  Remembering that the relativistic effects are suppressed in the presence of the Fermi
sea one can neglect the contribution from the antiquarks:
\begin{eqnarray}
\label{decompose}
f(t,\vec{r})=\sum_{s} u_s(\vec{q}) \alpha_{s}(\vec{q})\exp(-i Et+i\vec{q}\cdot\vec{r}) \\\nonumber 
g^\dagger (t,\vec{r})=\sum_{s} d^{\dagger}_s(-\vec{q})
\beta^{\ast}_s(\vec{q}) \exp(-i E t+i\vec{q}\cdot\vec{r}),
\end{eqnarray}
where $s$ describes the spin degrees of freedom, $\alpha_{s}(\vec{q})$ and $\beta_s(\vec{q})$ 
are some c-numbers. The energy $E$ describes the energy of the quark with respect to the
Fermi surface in the absence of the gap parameter $\Delta $.
The Dirac bispinors $u_s$ and $d_s$ satisfy the equations:
\begin{eqnarray}
(\vec{\alpha }\cdot\vec{q}+m\gamma_0-\mu)u_s(\vec{q}) = \epsilon_q u_s(\vec{q}) \\\nonumber
d^{\dagger}_s(-\vec{q}) (\vec{\alpha }\cdot\vec{q}-m\gamma_0+\mu) = d^{\dagger}_s(-\vec{q})\bar{\epsilon }_q,
\end{eqnarray}
and because momentum $\vec{q}$ can be in general the complex number one has to carefully distinguish between
the left and the right eigenvectors. Inserting (\ref{decompose})
into the equations (\ref{eom1}), using bispinors algebraic relations and
assuming constant value of the gap parameter one arrives at the Bogolubov - de Gennes equations
for $\alpha $ and $\beta $ parameters:
\begin{eqnarray}
\label{b-deG}
E \alpha_\uparrow (\vec{q}) = \epsilon_q \alpha_\uparrow (\vec{q})
+\Delta \beta^{\ast}_\downarrow (\vec{q}), \\\nonumber
E \beta^{\ast }_\downarrow (\vec{q}) = -\epsilon_q 
\beta^{\ast }_{\downarrow }(\vec{q})
+\Delta^\ast \alpha_{\uparrow }(\vec{q})
\end{eqnarray}
and the similar set of equations with reverse spin projection on the momentum $\vec{q}$:
\begin{eqnarray}
\label{b-deG2}
E \alpha_\downarrow (\vec{q}) = \epsilon_q \alpha_\downarrow (\vec{q})
- \Delta \beta^{\ast}_\uparrow (\vec{q}), \\\nonumber
E \beta^{\ast }_\uparrow (\vec{q}) = -\epsilon_q 
\beta^{\ast }_{\uparrow }(\vec{q})
-\Delta^\ast \alpha_{\downarrow }(\vec{q})
\end{eqnarray}
where $\epsilon_q = \sqrt{\vec{q}^2+m^2} - \mu = - \bar{\epsilon }_q$.
We have two possible solutions of the uniform equations (\ref{b-deG}) (the case of (\ref{b-deG2}) is similar):\\
for $\epsilon_q = \xi $
\begin{eqnarray}
\alpha_\uparrow = \sqrt{\frac{1}{2}\left( 1+\frac{\xi }{E}\right) }\exp{\left( i\frac{\delta }{2}\right) }\\\nonumber
\beta^{\ast }_\downarrow = \sqrt{\frac{1}{2}\left( 1-\frac{\xi }{E}\right) }\exp{\left( -i\frac{\delta }{2}\right) }
\end{eqnarray}
and for $\epsilon_q = -\xi $
\begin{eqnarray}
\alpha_\uparrow = \sqrt{\frac{1}{2}\left( 1-\frac{\xi }{E}\right) }\exp{\left( i\frac{\delta }{2}\right) }\\\nonumber
\beta^{\ast }_\downarrow = \sqrt{\frac{1}{2}\left( 1+\frac{\xi }{E}\right) }\exp{\left( -i\frac{\delta }{2}\right) }
\end{eqnarray}
where $\xi = \sqrt{E^2-|\Delta |^2}$ and $\delta $ is a phase of the gap. Let us point out that 
only for $E>|\Delta |$ the solution describes the propagating excitations. Finally the wavefunction describing
the quasiparticle excitation of energy $E$ in the 2SC phase takes the form:
\begin{eqnarray}
\label{quasi}
\psi (t,\vec{r})\equiv 
\left( \begin{array}{c}
f\\
g^\dagger
\end{array}\right) =
\left( \begin{array}{c}
\exp{\left( i\frac{\delta }{2}\right) }\sqrt{\frac{E+\xi }{2E}} 
\left( A u_\uparrow (\vec{q_1})-Bu_\downarrow (\vec{q_1})\right) \\
\exp{\left( -i\frac{\delta }{2}\right) }\sqrt{\frac{E-\xi }{2E}}\left( A d^{\dagger\,T}_\downarrow (-\vec{q_1})+
Bd^{\dagger\,T}_\uparrow (-\vec{q_1})\right)
\end{array}\right) \exp{(-i Et + i\vec{q_1}\cdot\vec{r})}+ \\\nonumber
\left( \begin{array}{c}
\exp{\left( i\frac{\delta }{2}\right) }\sqrt{\frac{E-\xi }{2E}} 
\left( C u_\uparrow (\vec{q_2})+Du_\downarrow (\vec{q_2})\right) \\
\exp{\left( -i\frac{\delta }{2}\right) }\sqrt{\frac{E+\xi }{2E}}\left( C d^{\dagger\,T}_\downarrow (-\vec{q_2})-
Dd^{\dagger\,T}_\uparrow (-\vec{q_2})\right)
\end{array}\right) \exp{(-i Et + i\vec{q_2}\cdot\vec{r})}
\end{eqnarray}
where $q_{1,2}=\pm\sqrt{(\mu\pm\xi )^2-m^2}$ and $A,B,C$ and $D$ are arbitrary constants.

\section{Andreev reflection}

Let us consider the simple physical problem of the quark scattering on the superconducting plane surface
placed at z=0 in space. If we approximate the boundary as the step function
($\Delta =0$ for $z<0$ and $\Delta = const$ for $z>0$) then we are looking for stationary solution of 
equations (\ref{eom1}), independently for negative and positive $z$, 
supplemented with the condition that matches the wavefunction at $z=0$.

Let us consider in detail the situation when the ``red'' quark, with energy $E$ and 
spin projection $\uparrow $ falls
at the plane boundary from the left. Then the wavefunctions (for $\Delta =0$) take the form:
\begin{equation}
\label{qgp}
\psi_< (t,z)=
\left( 
\begin{array}{c}
u_\uparrow (k)\exp{ikz}+Fu_\uparrow (-k)\exp{-ikz}\\
Hd^{\dagger\,T}_\downarrow (-p)\exp{ipz}
\end{array}
\right) \exp{\left( -iEt\right) }
\end{equation}
The coefficients
$F$ and $H$ describes the amplitude of the reflection of the $u$-particle 
and $d$-hole respectively. The first equation from (\ref{b-deG}) determine the value of 
the momentum of the $u$ quark as
$k^2=(\mu + E)^2-m^2$. In the similar fashion the second equation from (\ref{b-deG}) 
gives the value of the momentum of the reflected $d$ hole as $p^2=(\mu - E)^2-m^2$. 
For $z>0$ the quasiparticles excitation are described by the wavefunction 
$\psi_>(t,z)$ given by the expression (\ref{quasi}) with $B=D=0$ (no spin flip
in the scattering process). Let us note that for   
$0<E<|\Delta |$, the momenta of the quasiparticles has to be complex number:
\begin{equation}
q_{1,2} = \pm\sqrt{\mu^2-m^2-|\xi |^2 \pm 2i|\xi |\mu }
\end{equation} 
Particularly simple form of momenta one can find in the massless limit. In that
case $q_{1,2}=\pm \mu + i|\xi |$, where the signs are chosen as to
describe the vanishing of the wavefunction in 2SC phase for $z\rightarrow\infty $.
Now it is seen that for $E<|\Delta |$ in the 2SC phase the quasiparticle wavefunction is 
exponentially decaying with the suppression factor $|\xi | =\sqrt{|\Delta |^2-E^2}$.

The continuity conditions matching the wavefunctions are of the form:
\begin{eqnarray}
\label{cont}
\psi_<(t,z=0)=\psi_>(t,z=0)
\end{eqnarray}
Using this condition one can find the amplitudes of the scattering process:
\begin{eqnarray}
\label{ampl}
A=\sqrt{\frac{2E}{E+\xi }} \exp{\left( -i\frac{\delta }{2}\right) } + O\left(\frac{1}{\mu }\right)\\\nonumber
C=\sqrt{\frac{E-\xi }{2E}} \frac{mE}{\mu^2} \exp{\left( -i\frac{\delta }{2}\right) }+
O\left(\frac{1}{\mu^3}\right)\\\nonumber
F=\frac{m(E-\xi)}{\mu^2}+O\left(\frac{1}{\mu^3}\right)\\\nonumber
H=\sqrt{\frac{E-\xi }{E+\xi }}\exp{\left( i\delta\right) } + O\left(\frac{1}{\mu }\right)
\end{eqnarray}
In the limit where $|\Delta |,E,m<<\mu$ only $A$ and $H$ contribute which means that
only hole of the $d$-green quark can be reflected into the QGP phase and
the only quasiparticle with momentum $q_1$ can propagate in the 2SC phase.
It is interesting to point out that in the massless limit $F=C=0$ exactly. 
Additional insight we gain from the calculation of the
probability current. From the equations (\ref{eom1}) it follows that the probability current
$\vec{j}=\psi^{u\,\dagger }_R\vec{\alpha }\psi^{u}_R+
\psi^{d\,\dagger }_G\vec{\alpha }\psi^{d}_G$ is conserved. Using
(\ref{quasi},\ref{ampl}) one finds:   
\begin{equation}
\label{current}
j_z = 2\mu \left\{
\begin{array}{lcr}
0 & \mbox{for} & E<|\Delta |\\
\frac{2\xi }{E+\xi }+O\left(\frac{1}{\mu }\right) & \mbox{for} & E>|\Delta |
\end{array}
\right.
\end{equation}
The result (\ref{current}) has simple interpretation. If the quark $u$ have
energy below the gap it can not excite quasiparticle inside the 2SC phase.
In that case the matter waves does not propagate through the superconducting
medium. However if the $u$ quark possesses energy above the gap it
excites the quasiparticles with given transition coefficient. This result
may be of importance for the physics of the Neutron Stars.

\section{Conclusions} 

In this paper we describe the phenomena of Andreev reflection
at the junction between the cold quark-gluon plasma and
superconducting 2SC phase. We found that the hole of
different flavor and color than the incoming particle
is reflected toward the QGP phase. Inside the 2SC
phase the quasiparicles are excited, which for energy
$E$ of the incoming particle below the gap $|\Delta |$,
penetrate the 2SC phase for depth of the order of $1/\sqrt{|\Delta |^2-E^2}$
whereas for the case of $E>|\Delta |$ they propagate freely
inside the superconducting phase.
This result is similar to the situation one encounters in the condensed matter systems.
This can be expected because in the high density QCD
the purely relativistic phenomena are suppressed by
the powers of quark chemical potential.  

From the equation (\ref{current})
one can see that the transport phenomena (like heat transport or
density waves)
is strongly affected by the presence of the interfaces
inside the Neutron Stars. The suppression is exponential
of the form $\exp{(-|\Delta |/T)}$ where $T$ is a temperature
of the Neutron Star, usually much smaller than the
expected superconducting gap. This phenomena certainly
require more extensive analysis\footnote{Let us note that
probability current through the interface is not suppressed completely,
because in the 2SC phase we have unpaired blue quarks.}. 

The structure of the Neutron Stars can be complicated, thus
more work is needed for better understanding of the
Andreev reflection phenomena in superconducting QCD.
In particular one has to consider other possible 
superconducting phases like CFL or other possible
interfaces. Existence of such interfaces certainly
influences the dynamics of matter flow inside the
Neutron Stars.

\bigskip  

{\bf Acknowledgement} Let me specially thank Maciek A. Nowak for many important discussions.
Also discussions with Krishna Rajagopal, Sanjay Reddy and Motoi Tachibana were of
great value. This work was supported by Polish State Committee 
for Scientific Research, grant no. 2P 03B 094 19.    
  
\end{document}